# The Algonauts Project 2025 Challenge:
# How the Human Brain Makes Sense of Multimodal Movies

Alessandro T. Gifford[1,*] , Domenic Bersch[2] ,
Marie St-Laurent[3] , Basile Pinsard[3] ,
Julie Boyle[3,4] , Lune Bellec[3,4] ,
Aude Oliva[5] , Gemma Roig[2] , Radoslaw M. Cichy[1]

[1] Freie Universität Berlin, Berlin, Germany
[2] Goethe Universität Frankfurt, Frankfurt am Main, Germany
[3] Centre de recherche de l'institut universitaire de gériatrie de Montréal, Montréal, Canada
[4] Université de Montréal, Montréal, Canada
[5] Computer Science and Artificial Intelligence Laboratory, MIT, USA
* Correspondence: alessandro.gifford@gmail.com

**There is growing symbiosis between artificial and biological intelligence sciences: neural principles inspire new intelligent machines, which are in turn used to advance our theoretical understanding of the brain. To promote further collaboration between biological and artificial intelligence researchers, we introduce the 2025 edition of the Algonauts Project challenge:** *How the Human Brain Makes Sense of Multimodal Movies* **(https://algonautsproject.com/). In collaboration with the Courtois Project on Neuronal Modelling (CNeuroMod), this edition aims to bring forth a new generation of brain encoding models that are multimodal and that generalize well beyond their training distribution, by training them on the largest dataset of fMRI responses to movie watching available to date. Open to all, the 2025 challenge provides transparent, directly comparable results through a public leaderboard that is updated automatically after each submission to facilitate rapid model assessment and guide development. The challenge will end with a session at the 2025 Cognitive Computational Neuroscience (CCN) conference that will feature winning models. We welcome researchers interested in collaborating with the Algonauts Project by contributing ideas and datasets for future challenges.**

## Introduction

There is an ever growing symbiosis between research in biological and artificial intelligence. Artificial intelligence research produces algorithms that achieve human or superhuman performance, which are currently the best predictive models of neural responses to sensory stimulation (Cichy and Kaiser, 2019; Doerig et al., 2023; Kietzmann et al., 2019; Richards et al., 2022, 2019a; Saxe et al., 2021; Yamins and DiCarlo, 2016). The creation of increasingly powerful data-hungry models has pushed biological intelligence scientists to collect large-scale neural datasets that intensively sample single subjects (Allen et al., 2022; Gifford et al., 2022; Hebart et al., 2023; Kupers et al., 2024; Lahner et al., 2024; Naselaris et al., 2021). These large datasets facilitate the discovery and modeling of new neural mechanisms that in turn inspire the development of better artificial intelligence algorithms (Hassabis et al., 2017; Sinz et al., 2019; Ullman, 2019; Yang et al., 2022; Zador et al., 2023), and provide inductive biases to increase the robustness of artificial agents (Dapello et al., 2022; Guo et al., 2024; Li et al., 2019; Safarani et al., 2021; Shao et al., 2024; Toneva and Wehbe, 2019).

Neuroscience challenges and benchmarks catalyze the symbiosis between biological and artificial intelligences. They offer interactive platforms for scientists from different disciplines to cooperate and compete in building the best models of intelligence (Cichy et al., 2019a, 2019b, 2021; Gifford et al., 2023; Schrimpf et al., 2018, 2020; Turishcheva et al., 2023; Willeke et al., 2022a). Here, we introduce the 2025 edition of the Algonauts Project challenge, titled "*How the Human Brain Makes Sense of Multimodal Movies*". This edition is based on data from the Courtois Project on Neuronal Modelling (CNeuroMod, https://www.cneuromod.ca/), which has acquired the largest human dataset of functional magnetic resonance imaging (fMRI) neural responses to naturalistic multimodal stimulation to date (Boyle et al., 2023). The current challenge shares the goal of the three previous Algonauts Project challenges, which is to predict human brain responses through computational models (Cichy et al., 2021, 2019a; Gifford et al., 2023). Yet, it goes beyond the previous challenges in three important ways. First, to foster the development of computational models that better capture neural responses to real-life scenarios, the 2025 challenge uses multimodal naturalistic stimuli, which go beyond the unimodal images and short videos used in previous challenges (**Figure 1a**). Second, it provides an unprecedented amount of data on which to train and test computational models of the brain, with almost 80 hours of neural recordings per subject (**Figure 1b**). Third, to promote more robust models of intelligence, the 2025 challenge provides both in-distribution (ID) and out-of-distribution (OOD) tests of model performance, and the winning models will be selected solely based on their OOD performance (**Figure 1c**).

Our vision for the Algonauts Project 2025 challenge is to advance our understanding of the brain through more accurate and robust modeling of neural responses to multimodal naturalistic stimulation, and to improve the engineering of artificial models through biological intelligence constraints. By achieving both aims, we hope to strengthen the symbiosis between biological and artificial intelligence.

# Challenge Information

## Goal

Encoding models of neural responses are increasingly used as predictive and explanatory tools in computational neuroscience (Kay et al., 2008; Kell et al., 2018; Kriegeskorte and Douglas, 2019; Naselaris et al., 2011; Tuckute et al., 2023; Van Gerven, 2017; Wu et al., 2006; Yamins and DiCarlo, 2016). They consist of algorithms, typically based on deep learning architectures, that take stimuli as input, and output the corresponding neural activations, effectively modeling how the brain responds to (i.e., encodes) these stimuli. Thus, the goal of the 2025 challenge is to provide a platform for biological and artificial intelligence scientists to cooperate and compete in developing cutting-edge brain encoding models. Specifically, these models should predict the brain's response to multimodal naturalistic movies, and generalize outside their training distribution.

## Data

The Challenge data is based on the CNeuroMod dataset, which comprises extensive single-subject neural response samples collected during a range of controlled and naturalistic tasks and stimuli. The CNeuroMod dataset's unprecedented size, combined with the multimodal nature and diversity of its stimuli and tasks, makes it an ideal training and testing ground to build robust encoding models of fMRI responses to multimodal stimuli that generalize outside of their training distribution.

Challenge participants will train and evaluate their encoding models using a subset of CNeuroMod's data which includes almost 80 hours of multimodal movie stimuli and corresponding fMRI responses. Training data are from the CNeuroMod *Friends* dataset, for which subjects watched all episodes from the first six seasons of the sitcom *Friends* while undergoing fMRI, and from the CNeuroMod *Movie10* dataset for which they watched three feature films and a documentary. Evaluation data are from season seven of the Friends dataset, and from the viewing of out-of-distribution (OOD) movies.

The stimuli consist of movie visual frames, audio samples, and time-stamped language transcripts (**Figure 2a**). The neural data consist of whole-brain fMRI responses for four CNeuroMod subjects (sub-01, sub-02, sub-03 and sub-05), normalized to the Montreal Neurological Institute (MNI) spatial template (Brett et al., 2002), and processed as time series whose signal is assigned to 1,000 functionally defined brain parcels (Schaefer et al., 2018) (**Figure 2b**). Further information on the challenge stimuli and fMRI data is provided on the challenge data repository and development kit.

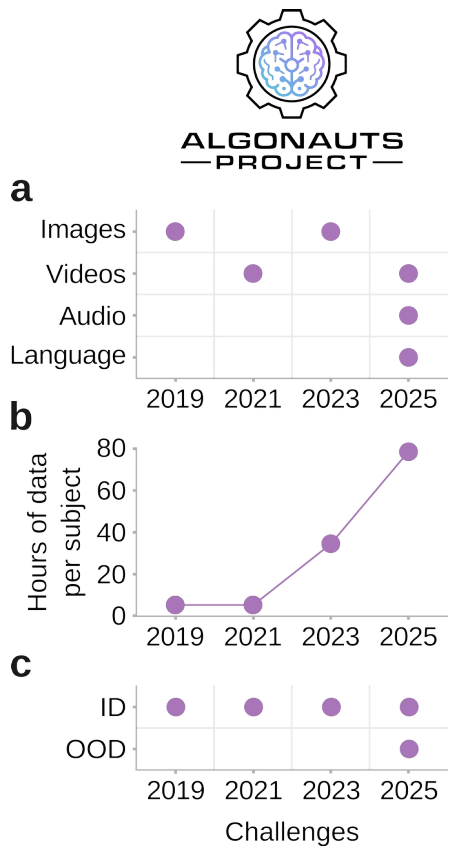

**Figure 1 | Comparison between editions of the Algonauts Project challenge (2019-2025).** **a**, Stimulus modalities. **b**, Hours of neural recordings per subject. **c**, Test set distributions with respect to the training distribution (ID = in-distribution; OOD = out-of-distribution).

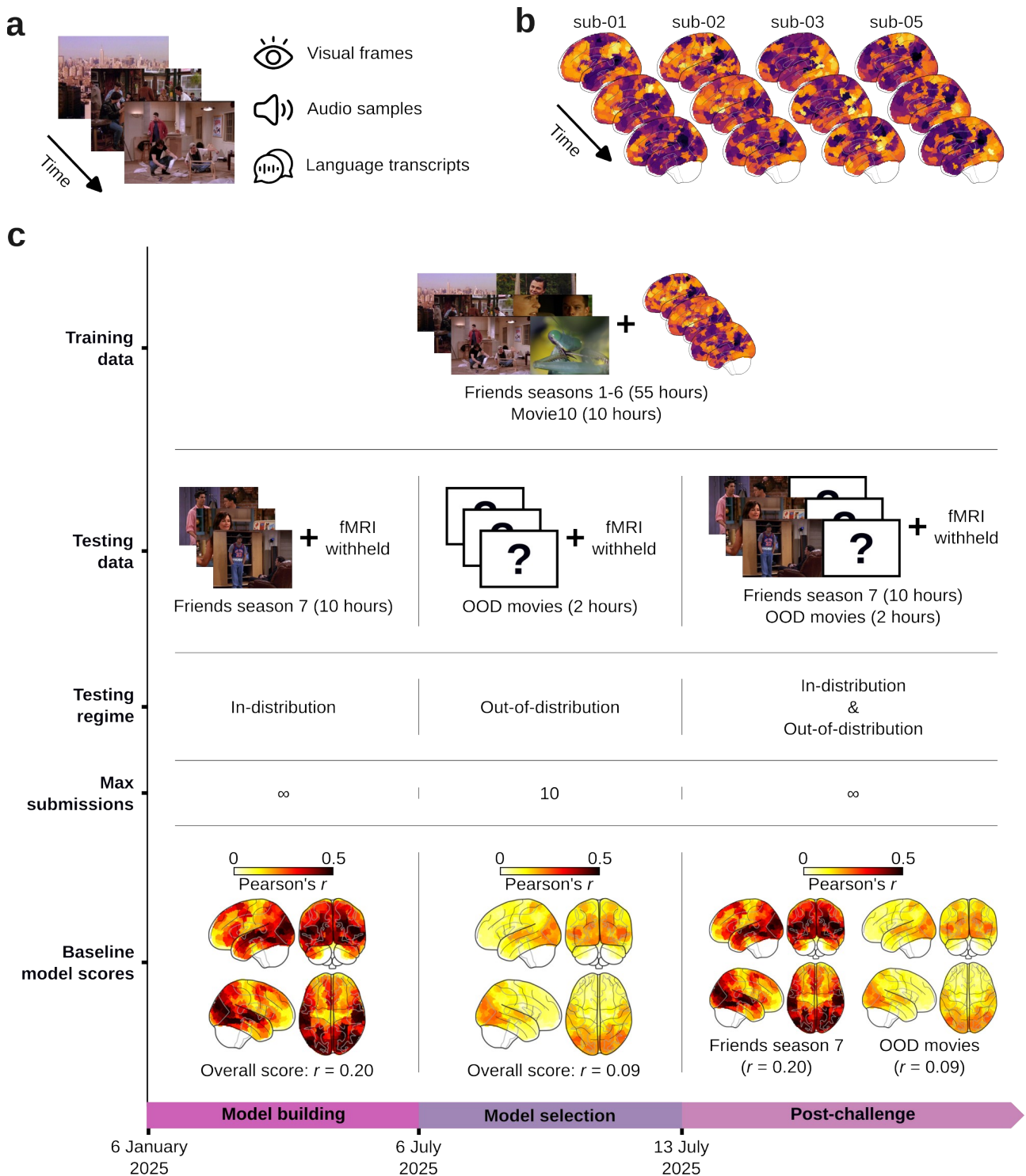

**Figure 2 | Challenge data and phases. a**, Multimodal movie stimuli, consisting of visual frames, audio samples, and time-stamped language transcripts. **b**, Whole-brain fMRI responses (time series) to the multimodal movies in four subjects. **c**, During the model building phase, models are trained using stimuli and corresponding fMRI responses for seasons 1 to 6 of the sitcom *Friends* and Movie10 (a set of four movies), and tested in-distribution (ID) on Friends season 7 (for which the fMRI responses are withheld) with unlimited submissions. During the model selection phase, the winning models are selected based on the accuracy of their predicted fMRI responses for out-of-distribution (OOD) movie stimuli (for which the fMRI responses are withheld) with up to ten submissions. The challenge will be followed by an indefinite post-challenge phase with unlimited submissions, which will serve as a public benchmark for both ID and OOD model validation.

## Phases

The challenge is hosted on [Codabench](), and consists of two main serial phases: a 6-months model building phase during which encoding models are trained and tested, followed by a 1-week model selection phase during which the winning models are selected based on their performance on a withheld OOD test set. To enforce strict tests of OOD generalization during the model selection phase, the two phases are based on data from different distributions, and have separate leaderboards. The challenge will be followed by an indefinite post-challenge phase, which will serve as a public benchmark for anyone wishing to test brain encoding models on multimodal movie data (**Figure 2c**).

## Model building phase (6 months)

During this first phase, challenge participants will train and test encoding models using movie stimuli and fMRI responses from the same distribution.

For model training, we provide 55 hours of movie stimuli and corresponding fMRI responses for each of the four subjects for all episodes of seasons 1 to 6 of the Friends dataset. We also provide 10 hours of movie stimuli and corresponding fMRI responses from the Movie10 dataset for which the same four subjects watched the following four movies: *The Bourne Supremacy*, *Hidden Figures*, *Life* (a BBC nature documentary), and *The Wolf of Wall Street*. Each movie was presented to each subject once, except for *Life* and *Hidden Figures* which were presented twice. Challenge participants can train their encoding models using these data.

For model testing, we provide 10 hours of movie stimuli for all episodes of seasons 7 of the Friends dataset, and withhold the corresponding fMRI responses for each subject. Challenge participants can test their encoding models against the withheld fMRI responses by submitting predicted fMRI responses for Friends season 7 to Codabench. After each submission, the scoring program will correlate (Pearson's *r*) the predicted fMRI responses for each parcel and subject with the recorded (withheld) fMRI responses across all Friends season 7 episodes, resulting in one correlation score for each parcel and subject. These correlation scores are averaged first across parcels and then across subjects, to obtain a single correlation score quantifying the performance of each submission.

## Model selection phase (1 week)

During this second phase, the winning models will be selected based on the accuracy of their predicted fMRI responses for withheld OOD movie stimuli.

We will provide 2 hours of OOD movie stimuli, and withhold the corresponding fMRI responses for each of the four subjects. The nature of the OOD movie stimuli will not be revealed until the beginning of the model selection phase. To participate in the winners selection process, challenge participants can submit their encoding models' predicted fMRI responses for the OOD movie stimuli to Codabench. After each submission, the scoring program will correlate the predicted fMRI responses for each parcel and subject with the recorded (withheld) fMRI responses, independently for each of the OOD movie stimuli, resulting in one correlation score for each parcel, OOD movie and subject. These correlation scores are averaged first across parcels, then across OOD movies, and finally across subjects, thus obtaining a single correlation score quantifying the performance of each submission.

## Post-challenge phase (indefinite)

Once the challenge is over, we will open an indefinite post-challenge phase which will serve as a public benchmark. This benchmark will consist of two separate leaderboards that will rank encoding models based on their fMRI predictions for ID (Friends season 7) or OOD (OOD movies) multimodal movie stimuli, respectively.

## Rules

To encourage broad participation, the challenge has a simple submission process with minimal rules.

Challenge participants can use any encoding model derived from any source and trained on any type of data. However, using recorded brain responses for Friends season 7 or the OOD movie stimuli is prohibited.

The winning models will be determined based on their performance in predicting fMRI responses for the OOD movie stimuli during the model selection phase.

Challenge participants can make an unlimited number of submissions during the model building phase, and a maximum of ten submissions during the model selection phase (**Figure 2c**) (the leaderboard of each phase is automatically updated after every submission). Each challenge participant can only compete using one account. Creating multiple accounts to increase the number of possible submissions will result in disqualification to the challenge.

To promote open science, challenge participants who wish to be considered for the winners selection will need to submit a short report to a preprint server describing their encoding algorithm. Furthermore, the top-3 performing teams are required to make their code openly available. Along with monetary prizes, they will be invited to present their encoding models during a talk at the [Cognitive Computational Neuroscience (CCN) conference](#) held in Amsterdam (Netherlands) in August 2025.

## Baseline model

The challenge baseline model score is computed with a linearizing encoding model (Naselaris et al., 2011) that extracts visual, audio, and language features from the movie stimuli, and linearly maps them onto fMRI responses. Its mean correlation scores over all parcels and subjects is $r = 0.20$ for Friends seasons 7 (model building phase), and $r = 0.09$ for the OOD movie stimuli (model selection phase) (**Figure 2c**).

### Development kit

To facilitate participation, we provide a [development kit](development kit) in Python which accompanies users through the challenge process, following four steps: (**i**) familiarizing with the challenge data; (**ii**) extracting the stimulus features used to train and validate an fMRI encoding model; (**iii**) training and validating an fMRI encoding model; (**iv**) preparing the predicted fMRI responses for the test stimuli in the correct format for submission to Codabench.

## Discussion

The Algonauts Project 2025 challenge advances the exploration of intelligence through a platform where scientists from different fields cooperate and compete in building computational models of the human brain. Just like past breakthroughs have resulted from mutual inspiration between diverse fields, multidisciplinary efforts will continue to help overcome unresolved problems in intelligence research. With this new edition of the challenge, we invite intelligence researchers to tackle one such unresolved problem at the forefront of biological and artificial intelligence research alike: the development of accurate and robust models of multisensory processing.

### Towards computational models of multimodal neural responses

Perception is a multisensory experience. For example, we see lips and mouths moving while hearing the sounds uttered by them, and we hear increasingly loud music as we see a mandolin player approaching us. Our brains promptly integrate these diverse sources of sensory information into one coherent percept through multimodal integration (Calvert et al., 2004; De Gelder and Bertelson, 2003; Ernst and Bülthoff, 2004). To assess brain responses in a more ecologically valid multimodal setting, the Algonauts Project 2025 challenge edition is, for the first time, inviting participants to model neural responses to visual, auditory and linguistic stimulation elicited during naturalistic movie watching. Our aim is to encourage the development of computational models of neural responses that incorporate diverse sources of sensory information present in life-like scenarios.

### Larger neural datasets innovate the interaction between biological and artificial intelligence research

During the last decade, deep learning algorithms have become state-of-the-art models of the brain (Cichy and Kaiser, 2019; Doerig et al., 2023; Kietzmann et al., 2019; Richards et al., 2019a; Saxe et al., 2021; Yamins and DiCarlo, 2016). A crucial factor for robust and successful modelling is the large amount of data on which they are trained (Russakovsky et al., 2015). We addressed this reality by basing the 2025 challenge on the unprecedented dataset size of almost 80 hours of fMRI responses for each of four subjects. This supports data-hungry modeling approaches, including the end-to-end training of deep learning architectures that predict neural responses to sensory stimuli (Allen et al., 2022; Gifford et al., 2022). Directly infusing deep learning models with brain data enables a novel type of interaction between biological and artificial intelligence, which in our opinion will catalyze breakthroughs in neuroscientific research.

### The importance of out-of-distribution generalization tests

Brain models are typically trained and tested on the same or similar data distributions, and as a result their out-of-distribution (OOD) generalization performance is unknown (Liu et al., 2021). This blind spot is an important limitation, because OOD tests benefit intelligence research in three significant ways. First, successful models should predict brain responses under a broad range of stimulations. In that sense, OOD generalization serves as a stricter but essential assessment of model robustness and validity than ID tests. Poor OOD generalization indicates that theoretical inferences of brain function might not apply beyond stimuli from the training distribution, suggesting that further development is needed to capture stimulus-to-brain-response relationships. Second, breaking down generalization scores across different types of OOD conditions can be informative by revealing stimulus properties that neural models fail to account for (Madan et al., 2024), hence providing explicit objectives of model improvement. Third, OOD generalization scores serve to distinguish among models that may otherwise have similar in-distribution test scores (Ren and Bashivan, 2024). Finding that one of several similarly-performing models generalizes better OOD can reveal important properties that make models successful (e.g., architecture, training diet, learning objective), informing the engineering of more robust models and inspiring new hypotheses about brain function. For these reasons, the 2025 challenge provides testing stimuli both inside and outside the training data distribution, and will select winning models solely based on their OOD generalization performance.

### The explanatory relevance of prediction

While prediction and explanation are two related goals in science, one does not imply the other (Shmueli, 2010; Yarkoni and Westfall, 2017). Better predictive models of neural responses, by themselves, do not necessarily contribute to a better explanatory understanding of the brain. However, better prediction mediates better explanations in several crucial ways.

First, adequate explanations should generate successful predictions. Thus, predictive modeling provides a reality check to validate existing theories (Shmueli, 2010), for example by adjudicating between competing interpretations, hypotheses or models based on their predictive power (Golan et al., 2020).

Second, model ranking and comparison based on prediction accuracy sheds light on properties that make models successful, which in turn helps generate new explanatory hypotheses regarding brain mechanisms. This is particularly relevant for large and rich neural datasets (Allen et al., 2022; Gifford et al., 2022; Hebart et al., 2023; Lahner et al., 2024) from which complex relationships can be extracted that can be hard to formalize based on pure theoretical grounds (Shmueli, 2010).

Third, predictive models enable the quick and cheap generation of massive amounts of denoised in silico neural response estimates to any amount and type of input stimuli, thereby allowing for unprecedentedly large upscaling of the data space on which to explore and test hypotheses of brain function. Novel findings on in silico neural responses are then empirically validated in vivo to achieve explanatory understanding (Gifford et al., 2024; Jain et al., 2024; Mathis et al., 2024).

## Relation to similar initiatives

The Algonauts Project shares similarities with initiatives such as Brain-Score (Schrimpf et al., 2020, 2018) and the Sensorium competition (Turishcheva et al., 2023; Willeke et al., 2022b), which also establish benchmarks and leaderboards for computational models of the brain. However, the Algonauts Project 2025 challenge differs from these other efforts by relying on whole brain fMRI data from the dataset that most intensively samples single human subjects, by focusing on naturalistic multimodal stimulation, and by premiating models that best generalize out-of-distribution.

## The future of the Algonauts Project

We hope that the 2025 edition of the Algonauts Project will continue to inspire new challenges and collaborations that bring together artificial and biological intelligence science, as both communities can benefit from jointly tackling open questions on how perception and cognition are solved in brains and machines (Doerig et al., 2023; Hassabis et al., 2017; Richards et al., 2022, 2019b; Zador et al., 2023). As an example, to promote models that predict a wider range of cognitive mechanisms, and thus better capture the complex dynamics of real brains, an interesting future avenue is to model neural responses beyond passive stimulation tasks. This feat could be achieved by training models with neural datasets that capture higher level cognitive processes such as attention, goal seeking, decision making and visuomotor integration, for instance with datasets sampling neural responses during video game playing. We strongly welcome researchers interested in initiating similar initiatives, or in collaborating with the Algonauts Project by contributing ideas and datasets for future challenges.

## Acknowledgments

R.M.C. is supported by German Research Council (DFG) grants (CI 241/1-3, CI 241/1-7, INST 272/297-1) and the European Research Council (ERC) starting grant (ERC-StG-2018-803370). G.R. is supported by the German Research Foundation (DFG Research Unit FOR 5368 ARENA). The Courtois project on neural modelling was made possible by a generous donation from the Courtois foundation, administered by the Fondation Institut Gériatrie Montréal at CIUSSS du Centre-Sud-de-l'île-de-Montréal and University of Montreal. The CNeuroMod data used in the Algonauts 2025 Challenge has been openly shared under a Creative Commons CC0 license by a subset of CNeuroMod participants through the Canadian Open Neuroscience Platform (CONP), funded by Brain Canada and based at McGill University, Canada.

## References

Allen, E.J., St-Yves, G., Wu, Y., Breedlove, J.L., Prince, J.S., Dowdle, L.T., Nau, M., Caron, B., Pestilli, F., Charest, I., Hutchinson, J.B., Naselaris, T., Kay, K., 2022. A massive 7T fMRI dataset to bridge cognitive neuroscience and artificial intelligence. Nat. Neurosci. 25, 116–126. https://doi.org/10.1038/s41593-021-00962-x

Boyle, J., Pinsard, B., Borghesani, V., Paugam, F., DuPre, E., Bellec, P., 2023. The Courtois NeuroMod project: quality assessment of the initial data release (2020), in: 2023 Conference on Cognitive Computational Neuroscience. Presented at the 2023 Conference on Cognitive Computational Neuroscience, Cognitive Computational Neuroscience, Oxford, UK. https://doi.org/10.32470/CCN.2023.1602-0

Brett, M., Johnsrude, I.S., Owen, A.M., 2002. The problem of functional localization in the human brain. Nat. Rev. Neurosci. 3, 243–249. https://doi.org/10.1038/nrn756

Calvert, G., Spence, C., Stein, B.E. (Eds.), 2004. The handbook of multisensory processes. MIT Press, Cambridge, Mass.

Cichy, R.M., Dwivedi, K., Lahner, B., Lascelles, A., Iamshchinina, P., Graumann, M., Andonian, A., Murty, N.A.R., Kay, K., Roig, G., Oliva, A., 2021. The Algonauts Project 2021 Challenge: How the Human Brain Makes Sense of a World in Motion. https://doi.org/10.48550/ARXIV.2104.13714

Cichy, R.M., Kaiser, D., 2019. Deep Neural Networks as Scientific Models. Trends Cogn. Sci. 23, 305–317. https://doi.org/10.1016/j.tics.2019.01.009

Cichy, R.M., Roig, G., Andonian, A., Dwivedi, K., Lahner, B., Lascelles, A., Mohsenzadeh, Y., Ramakrishnan, K., Oliva, A., 2019a. The Algonauts Project: A Platform for Communication between the Sciences of Biological and Artificial Intelligence. https://doi.org/10.48550/ARXIV.1905.05675

Cichy, R.M., Roig, G., Oliva, A., 2019b. The Algonauts Project. Nat. Mach. Intell. 1, 613–613.

https://doi.org/10.1038/s42256-019-0127-z

Dapello, J., Kar, K., Schrimpf, M., Geary, R., Ferguson, M., Cox, D.D., DiCarlo, J.J., 2022. Aligning Model and Macaque Inferior Temporal Cortex Representations Improves Model-to-Human Behavioral Alignment and Adversarial Robustness. https://doi.org/10.1101/2022.07.01.498495

De Gelder, B., Bertelson, P., 2003. Multisensory integration, perception and ecological validity. Trends Cogn. Sci. 7, 460–467. https://doi.org/10.1016/j.tics.2003.08.014

Doerig, A., Sommers, R.P., Seeliger, K., Richards, B., Ismael, J., Lindsay, G.W., Kording, K.P., Konkle, T., Van Gerven, M.A.J., Kriegeskorte, N., Kietzmann, T.C., 2023. The neuroconnectionist research programme. Nat. Rev. Neurosci. 24, 431–450. https://doi.org/10.1038/s41583-023-00705-w

Ernst, M.O., Bülthoff, H.H., 2004. Merging the senses into a robust percept. Trends Cogn. Sci. 8, 162–169. https://doi.org/10.1016/j.tics.2004.02.002

Gifford, A.T., Dwivedi, K., Roig, G., Cichy, R.M., 2022. A large and rich EEG dataset for modeling human visual object recognition. NeuroImage 264, 119754. https://doi.org/10.1016/j.neuroimage.2022.119754

Gifford, A.T., Jastrzębowska, M.A., Singer, J.J.D., Cichy, R.M., 2024. In silico discovery of representational relationships across visual cortex. https://doi.org/10.48550/ARXIV.2411.10872

Gifford, A.T., Lahner, B., Saba-Sadiya, S., Vilas, M.G., Lascelles, A., Oliva, A., Kay, K., Roig, G., Cichy, R.M., 2023. The Algonauts Project 2023 Challenge: How the Human Brain Makes Sense of Natural Scenes. https://doi.org/10.48550/ARXIV.2301.03198

Golan, T., Raju, P.C., Kriegeskorte, N., 2020. Controversial stimuli: Pitting neural networks against each other as models of human cognition. Proc. Natl. Acad. Sci. 117, 29330–29337. https://doi.org/10.1073/pnas.1912334117

Guo, M., Choksi, B., Sadiya, S., Gifford, A.T., Vilas, M.G., Cichy, R.M., Roig, G., 2024. Limited but consistent gains in adversarial robustness by co-training object recognition models with human EEG. https://doi.org/10.48550/ARXIV.2409.03646

Hassabis, D., Kumaran, D., Summerfield, C., Botvinick, M., 2017. Neuroscience-Inspired Artificial Intelligence. Neuron 95, 245–258. https://doi.org/10.1016/j.neuron.2017.06.011

Hebart, M.N., Contier, O., Teichmann, L., Rockter, A.H., Zheng, C.Y., Kidder, A., Corriveau, A., Vaziri-Pashkam, M., Baker, C.I., 2023. THINGS-data, a multimodal collection of large-scale datasets for investigating object representations in human brain and behavior. eLife 12, e82580. https://doi.org/10.7554/eLife.82580

Jain, S., Vo, V.A., Wehbe, L., Huth, A.G., 2024. Computational Language Modeling and the Promise of In Silico Experimentation. Neurobiol. Lang. 5, 80–106. https://doi.org/10.1162/nol_a_00101

Kay, K.N., Naselaris, T., Prenger, R.J., Gallant, J.L., 2008. Identifying natural images from human brain activity. Nature 452, 352–355. https://doi.org/10.1038/nature06713

Kell, A.J.E., Yamins, D.L.K., Shook, E.N., Norman-Haignere, S.V., McDermott, J.H., 2018. A Task-Optimized Neural Network Replicates Human Auditory Behavior, Predicts Brain Responses, and Reveals a Cortical Processing Hierarchy. Neuron 98, 630-644.e16. https://doi.org/10.1016/j.neuron.2018.03.044

Kietzmann, T.C., McClure, P., Kriegeskorte, N., 2019. Deep Neural Networks in Computational Neuroscience, in: Oxford Research Encyclopedia of Neuroscience. Oxford University Press. https://doi.org/10.1093/acrefore/9780190264086.013.46

Kriegeskorte, N., Douglas, P.K., 2019. Interpreting encoding and decoding models. Curr. Opin. Neurobiol. 55, 167–179. https://doi.org/10.1016/j.conb.2019.04.002

Kupers, E.R., Knapen, T., Merriam, E.P., Kay, K.N., 2024. Principles of intensive human neuroimaging. Trends Neurosci. S0166223624001838. https://doi.org/10.1016/j.tins.2024.09.011

Lahner, B., Dwivedi, K., Iamshchinina, P., Graumann, M., Lascelles, A., Roig, G., Gifford, A.T., Pan, B., Jin, S., Ratan Murty, N.A., Kay, K., Oliva, A., Cichy, R., 2024. Modeling short visual events through the BOLD moments video fMRI dataset and metadata. Nat. Commun. 15, 6241. https://doi.org/10.1038/s41467-024-50310-3

Li, Z., Brendel, W., Walker, E.Y., Cobos, E., Muhammad, T., Reimer, J., Bethge, M., Sinz, F.H., Pitkow, X., Tolias, A.S., 2019. Learning From Brains How to Regularize Machines. https://doi.org/10.48550/ARXIV.1911.05072

Liu, J., Shen, Z., He, Y., Zhang, X., Xu, R., Yu, H., Cui, P., 2021. Towards Out-Of-Distribution Generalization: A Survey. https://doi.org/10.48550/ARXIV.2108.13624

Madan, S., Xiao, W., Cao, M., Pfister, H., Livingstone, M., Kreiman, G., 2024. Benchmarking Out-of-Distribution Generalization Capabilities of DNN-based Encoding Models for the Ventral Visual Cortex. https://doi.org/10.48550/ARXIV.2406.16935

Mathis, M.W., Perez Rotondo, A., Chang, E.F., Tolias, A.S., Mathis, A., 2024. Decoding the brain: From neural representations to mechanistic models. Cell 187, 5814–5832. https://doi.org/10.1016/j.cell.2024.08.051

Naselaris, T., Allen, E., Kay, K., 2021. Extensive sampling for complete models of individual brains. Curr. Opin. Behav. Sci. 40, 45–51. https://doi.org/10.1016/j.cobeha.2020.12.008

Naselaris, T., Kay, K.N., Nishimoto, S., Gallant, J.L., 2011. Encoding and decoding in fMRI. NeuroImage 56, 400–410. https://doi.org/10.1016/j.neuroimage.2010.07.073

Ren, Y., Bashivan, P., 2024. How well do models of visual cortex generalize to out of distribution samples? PLOS Comput. Biol. 20, e1011145. https://doi.org/10.1371/journal.pcbi.1011145

Richards, B., Tsao, D., Zador, A., 2022. The application of artificial intelligence to biology and neuroscience. Cell 185, 2640–2643. https://doi.org/10.1016/j.cell.2022.06.047

Richards, B.A., Lillicrap, T.P., Beaudoin, P., Bengio, Y., Bogacz, R., Christensen, A., Clopath, C., Costa, R.P., De Berker, A., Ganguli, S., Gillon, C.J., Hafner, D., Kepecs, A., Kriegeskorte, N., Latham, P., Lindsay, G.W., Miller, K.D., Naud, R., Pack, C.C., Poirazi, P., Roelfsema, P., Sacramento, J., Saxe, A., Scellier, B., Schapiro, A.C., Senn, W., Wayne, G., Yamins, D., Zenke, F., Zylberberg, J., Therien, D., Kording, K.P., 2019a. A deep learning framework for neuroscience. Nat. Neurosci. 22, 1761–1770. https://doi.org/10.1038/s41593-019-0520-2

Richards, B.A., Lillicrap, T.P., Beaudoin, P., Bengio, Y., Bogacz, R., Christensen, A., Clopath, C., Costa, R.P., De Berker, A., Ganguli, S., Gillon, C.J., Hafner, D., Kepecs, A., Kriegeskorte, N., Latham, P., Lindsay, G.W., Miller, K.D., Naud, R., Pack, C.C.,

Poirazi, P., Roelfsema, P., Sacramento, J., Saxe, A., Scellier, B., Schapiro, A.C., Senn, W., Wayne, G., Yamins, D., Zenke, F., Zylberberg, J., Therien, D., Kording, K.P., 2019b. A deep learning framework for neuroscience. Nat. Neurosci. 22, 1761–1770. https://doi.org/10.1038/s41593-019-0520-2

Russakovsky, O., Deng, J., Su, H., Krause, J., Satheesh, S., Ma, S., Huang, Z., Karpathy, A., Khosla, A., Bernstein, M., Berg, A.C., Fei-Fei, L., 2015. ImageNet Large Scale Visual Recognition Challenge. Int. J. Comput. Vis. 115, 211–252. https://doi.org/10.1007/s11263-015-0816-y

Safarani, S., Nix, A., Willeke, K., Cadena, S.A., Restivo, K., Denfield, G., Tolias, A.S., Sinz, F.H., 2021. Towards robust vision by multi-task learning on monkey visual cortex. https://doi.org/10.48550/ARXIV.2107.14344

Saxe, A., Nelli, S., Summerfield, C., 2021. If deep learning is the answer, what is the question? Nat. Rev. Neurosci. 22, 55–67. https://doi.org/10.1038/s41583-020-00395-8

Schaefer, A., Kong, R., Gordon, E.M., Laumann, T.O., Zuo, X.-N., Holmes, A.J., Eickhoff, S.B., Yeo, B.T.T., 2018. Local-Global Parcellation of the Human Cerebral Cortex from Intrinsic Functional Connectivity MRI. Cereb. Cortex 28, 3095–3114. https://doi.org/10.1093/cercor/bhx179

Schrimpf, M., Kubilius, J., Hong, H., Majaj, N.J., Rajalingham, R., Issa, E.B., Kar, K., Bashivan, P., Prescott-Roy, J., Geiger, F., Schmidt, K., Yamins, D.L.K., DiCarlo, J.J., 2018. Brain-Score: Which Artificial Neural Network for Object Recognition is most Brain-Like? https://doi.org/10.1101/407007

Schrimpf, M., Kubilius, J., Lee, M.J., Ratan Murty, N.A., Ajemian, R., DiCarlo, J.J., 2020. Integrative Benchmarking to Advance Neurally Mechanistic Models of Human Intelligence. Neuron 108, 413–423. https://doi.org/10.1016/j.neuron.2020.07.040

Shao, Z., Ma, L., Li, B., Beck, D.M., 2024. Leveraging the Human Ventral Visual Stream to Improve Neural Network Robustness. https://doi.org/10.48550/ARXIV.2405.02564

Shmueli, G., 2010. To Explain or to Predict? Stat. Sci. 25. https://doi.org/10.1214/10-STS330

Sinz, F.H., Pitkow, X., Reimer, J., Bethge, M., Tolias, A.S., 2019. Engineering a Less Artificial Intelligence. Neuron 103, 967–979. https://doi.org/10.1016/j.neuron.2019.08.034

Toneva, M., Wehbe, L., 2019. Interpreting and improving natural-language processing (in machines) with natural language-processing (in the brain). https://doi.org/10.48550/ARXIV.1905.11833

Tuckute, G., Feather, J., Boebinger, D., McDermott, J.H., 2023. Many but not all deep neural network audio models capture brain responses and exhibit correspondence between model stages and brain regions. PLOS Biol. 21, e3002366. https://doi.org/10.1371/journal.pbio.3002366

Turishcheva, P., Fahey, P.G., Hansel, L., Froebe, R., Ponder, K., Vystrčilová, M., Willeke, K.F., Bashiri, M., Wang, E., Ding, Z., Tolias, A.S., Sinz, F.H., Ecker, A.S., 2023. The Dynamic Sensorium competition for predicting large-scale mouse visual cortex activity from videos. https://doi.org/10.48550/ARXIV.2305.19654

Ullman, S., 2019. Using neuroscience to develop artificial intelligence. Science 363, 692–693. https://doi.org/10.1126/science.aau6595

Van Gerven, M.A.J., 2017. A primer on encoding models in sensory neuroscience. J. Math. Psychol. 76, 172–183. https://doi.org/10.1016/j.jmp.2016.06.009

Willeke, K.F., Fahey, P.G., Bashiri, M., Pede, L., Burg, M.F., Blessing, C., Cadena, S.A., Ding, Z., Lurz, K.-K., Ponder, K., Muhammad, T., Patel, S.S., Ecker, A.S., Tolias, A.S., Sinz, F.H., 2022a. The Sensorium competition on predicting large-scale mouse primary visual cortex activity. https://doi.org/10.48550/ARXIV.2206.08666

Willeke, K.F., Fahey, P.G., Bashiri, M., Pede, L., Burg, M.F., Blessing, C., Cadena, S.A., Ding, Z., Lurz, K.-K., Ponder, K., Muhammad, T., Patel, S.S., Ecker, A.S., Tolias, A.S., Sinz, F.H., 2022b. The Sensorium competition on predicting large-scale mouse primary visual cortex activity. https://doi.org/10.48550/ARXIV.2206.08666

Wu, M.C.-K., David, S.V., Gallant, J.L., 2006. Complete Functional Characterization Of Sensory Neurons By System Identification. Annu. Rev. Neurosci. 29, 477–505. https://doi.org/10.1146/annurev.neuro.29.051605.113024

Yamins, D.L.K., DiCarlo, J.J., 2016. Using goal-driven deep learning models to understand sensory cortex. Nat. Neurosci. 19, 356–365. https://doi.org/10.1038/nn.4244

Yang, X., Yan, J., Wang, W., Li, S., Hu, B., Lin, J., 2022. Brain-inspired models for visual object recognition: an overview. Artif. Intell. Rev. 55, 5263–5311. https://doi.org/10.1007/s10462-021-10130-z

Yarkoni, T., Westfall, J., 2017. Choosing Prediction Over Explanation in Psychology: Lessons From Machine Learning. Perspect. Psychol. Sci. 12, 1100–1122. https://doi.org/10.1177/1745691617693393

Zador, A., Escola, S., Richards, B., Ölveczky, B., Bengio, Y., Boahen, K., Botvinick, M., Chklovskii, D., Churchland, A., Clopath, C., DiCarlo, J., Ganguli, S., Hawkins, J., Körding, K., Koulakov, A., LeCun, Y., Lillicrap, T., Marblestone, A., Olshausen, B., Pouget, A., Savin, C., Sejnowski, T., Simoncelli, E., Solla, S., Sussillo, D., Tolias, A.S., Tsao, D., 2023. Catalyzing next-generation Artificial Intelligence through NeuroAI. Nat. Commun. 14, 1597. https://doi.org/10.1038/s41467-023-37180-x